\documentclass[12pt]{iopart}

\usepackage{graphicx}
\usepackage{iopams}
\usepackage{amssymb}

  \expandafter\let\csname equation*\endcsname\relax
  \expandafter\let\csname endequation*\endcsname\relax
\usepackage{amsmath}

\newcommand{\ket}[1]{\left|{#1}\right\rangle}
\newcommand{\bra}[1]{\left\langle{#1}\right|}

\newcommand{\ketbra}[2]{\left|{#1}\right\rangle\left\langle{#2}\right|}
\newcommand{\trb}[1]{\textrm{tr}\left\{#1\right\}}
\newcommand{\ptrb}[2]{\textrm{tr}_{#1}\left\{#2\right\}}
\newcommand{\beq}{\begin{equation}}
\newcommand{\eeq}{\end{equation}}

\usepackage{latexsym}
\usepackage{hyperref}
\usepackage{epsfig}
\usepackage{graphics}
\usepackage{xspace}

\begin{document}

\title[Reconstruction of photon number conditioned states]{Reconstruction of photon number conditioned states
  using phase randomised homodyne measurements}

\author{H.~M.~Chrzanowski$^1$, S.~M.~Assad$^1$, J.~Bernu$^1$, B.~Hage$^{2,1}$, A.~P.~Lund$^3$, T.~C.~Ralph$^3$, P.~K.~Lam$^1$ and T.~Symul$^1$}

\address{$^1$ Centre for Quantum Computation and Communication
Technology, Department of Quantum Science, Research School of Physics
and Engineering, Australian National University, Canberra ACT 0200,
Australia.}

\address{$^2$ Institut f\"{u}r Physik, Universit\"{a}t Rostock, 18055
Rostock, Germany.}

\address{$^3$ Centre for Quantum Computation and Communication
Technology, Department of Physics, University of Queensland, St. Lucia
QLD 4072, Australia.}
\ead{helen.chrzanowski@anu.edu.au}

\begin{abstract}
 We experimentally demonstrate the reconstruction of a photon number
 conditioned state without using a photon number discriminating
 detector. By using only phase randomised homodyne measurements, we
 reconstruct up to the three photon subtracted squeezed vacuum
 state. The reconstructed Wigner functions of these states show
 regions of pronounced negativity, signifying the non-classical nature
 of the reconstructed states. The techniques presented allow for
 complete characterisation of the role of a conditional measurement on
 an ensemble of states, and might prove useful in systems where photon
 counting is still technically challenging.
\end{abstract}


\section{Introduction}

Central to the weirdness of quantum mechanics is the notion of
wave-particle duality, where classical concepts of particle or wave
behaviour alone cannot provide a complete description of quantum
objects. When investigating quantum systems, information concerning
one description is typically sacrificed in favour of the other,
depending on which description suits your endeavour. Probing the
continuous variables of an infinite Hilbert space, such as the
amplitude and phase of a light field, is often viewed as less
interesting than probing the quantised variables of a quantum
system. This is largely due to the fact that, given current
technology, when probing the continuous variables (CV) of a quantum
system alone, one is restricted to transformations that map Gaussian
states onto Gaussian states.  Nevertheless, the idea of measuring the
quantised nature of light with only CV techniques has been
theoretically~\cite{Leonhardt:1995ud,Richter:1998bk,Ralph:2000eo,Ralph:2008kq} and
experimentally~\cite{Banaszek:1997vv,Vasilyev:2000tu,Webb:2006fy,Grosse:2007fu}
investigated.

The usual CV toolbox of Gaussian transformations, comprising beam
splitters, displacements, rotations, squeezing, homodyne and
heterodyne detection allows for deterministic manipulation of quantum
optical states that can be experimentally realised with typically
very high efficiency. However, the absence of a strong non-linearity
within this toolbox severely handicaps the reach of CV techniques for
quantum information processing applications \cite{Eisert:2002wb,Giedke:2002wi}.
Conversely, DV is implicitly non-linear---forgoing
determinism to harness the measurement-induced non-linearity of a
photon-counting measurement. Recently, there has been a move to
hybridise both CV and DV techniques for quantum information purposes,
as one non-Gaussian operation, when combined with Gaussian resources
and operations, is sufficient to realise universal quantum computing
\cite{VanLoock:2011wn}.

Here we present the CV analog of the photon counting measurement,
whereby we replace a non-deterministic photon counting measurement
with a deterministic phase randomised measurement of the field
quadratures. This extends the ideas reported in
\cite{Ralph:2008kq,Chrzanowski:2011cl} to show how the requirement of
a photon counting measurement can be replaced by CV measurements for
the reconstruction of the statistics of non-Gaussian states. This
approach forgoes the shot by shot nature of DV photon counting in
favour of ensemble measurements, and consequently cannot be
appropriated for state preparation. As we only preform Gaussian
measurements, all the directly measured statistics remain Gaussian and
the `non-Gaussianity' emerges in the nature of post-processing
preformed. The inherently ensemble nature of the technique and our
restriction to Gaussian measurements ensure it can never be used to
prepare a non-Gaussian state---in accordance with the limitations of
Gaussian toolbox. It does, however, still permit access to the same
non-Gaussian statistics that were previously only accessible with the
requirement of a projective photon counting measurement. Using this
method, we have successfully reconstructed the non-Gaussian 1, 2 and 3
photon subtracted squeezed vacuum (PSSV) states.

The context for the implementation of this protocol will be the
characterisation of the PSSV states. Also coined `kitten states', due
to their high fidelity with small amplitude Schr\"{o}dinger cat states, these states are typically prepared by
annihilating one or more photons from a squeezed vacuum state
\cite{Dakna:1997ur}. This `annihilation' is experimentally realised
to high fidelity with a beam splitter of weak reflectivity and a
conditional photon counting measurement, such that the detection of a
photon in the reflected mode heralds the successful subtraction of a
photon. Interest in such states was mainly prompted by optical quantum
computing \cite{Gilchrist:2004te,Lund:2008hr}, but they are also of interest for
metrology and entanglement distillation
\cite{Opatrny:2000vp,Ourjoumtsev:2007he}. Experiments involving the
generation of kitten states were amongst the first hybridisation
experiments---bridging the gap between two historically distinct areas
of quantum optics
\cite{Ourjoumtsev:2007va,Ourjoumtsev:2006vb,NeergaardNielsen:2006wm,Wakui:07,Gerrits:2010tk}.

This paper is organised as follows: in Section 2, we discuss the
theory linking homodyne detection and photon counting
measurement. Section 3 discusses the experimental implementation. We present
the experimental results in Section 4. The Appendix provides the
conditioning functions that transform photon
counting measurements to homodyne observables.

\section{Theory}
We want to design the homodyne equivalent of a heralded photon discriminating 
measurement. The setup consist of a correlated two mode state $\rho_{ab}$, where
mode $a$ is used to condition the outcome of mode $b$ (see Figure \ref{figure2}).
The conditioning measurement consists of sampling the homodyne
observable $\hat X_a^{\phi}$ in a phase randomised manner such that each
quadrature angle, $\phi$ contributes equally. Here, $\hat{X}^{\phi}_a=e^{-i \phi}\hat{a}_a+e^{i
  \phi}\hat{a}_a^\dagger$, where $\hat{a}_a$ and $\hat{a}_a^\dagger$
are the annihilation and creation operators in mode $a$ and $\phi$ is
the field quadrature angle. The conditioned mode $b$ is then
characterised via homodyne tomography.

\begin{figure*}
\includegraphics[width=15cm]{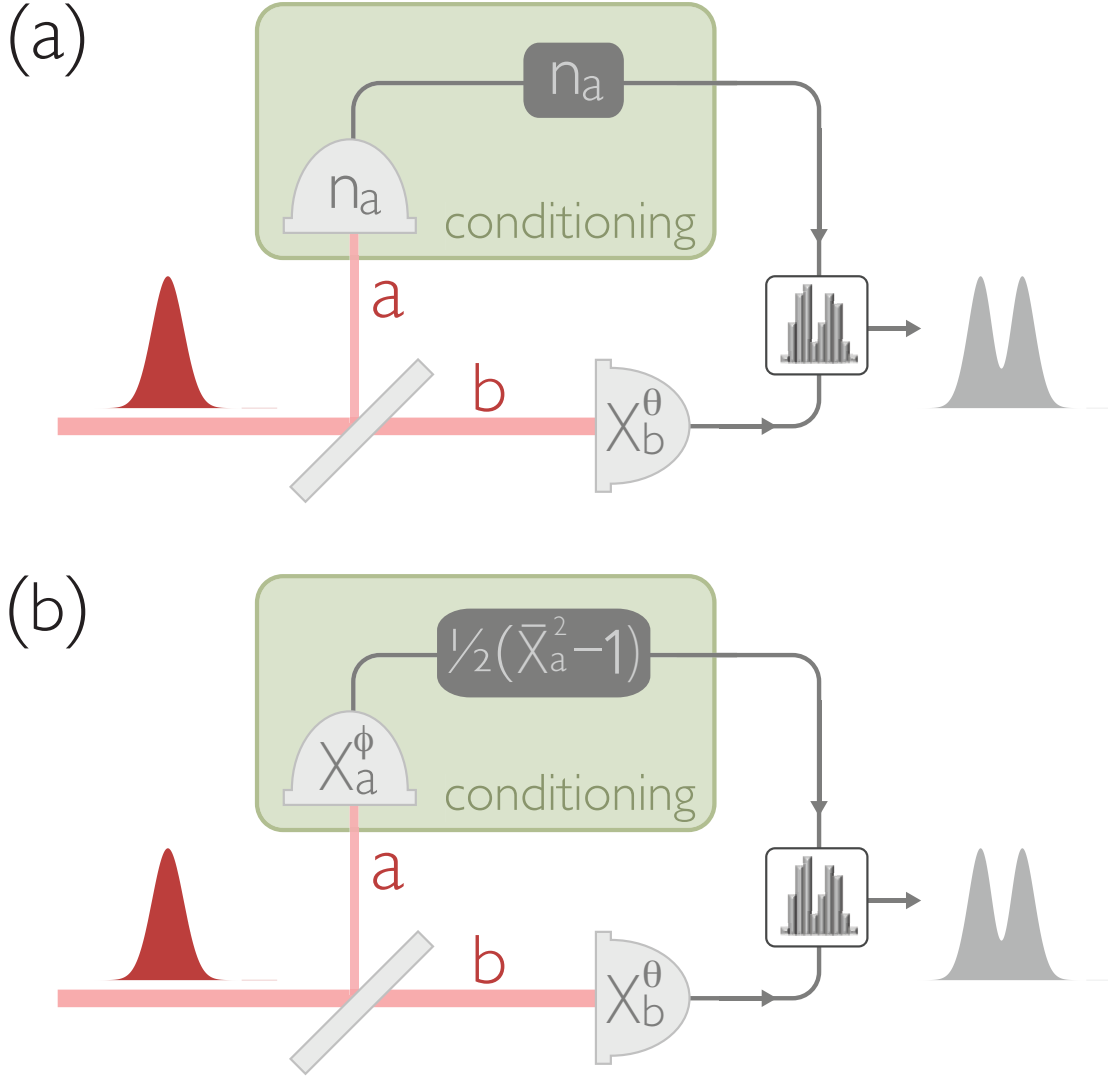}
\caption{(a) A photon number discriminating detector heralds a successful preparation. The correctly prepared state is subsequently reconstructed via homodyne tomography.  (b) The same statistics of the state heralded by a photon counting measurement can be retrieved by replacing the photon number discriminating detector with a phase randomised homodyne detection and appropriate post-processing.}
\label{fig:con}
\end{figure*}

\begin{figure*}
\includegraphics[width=15cm]{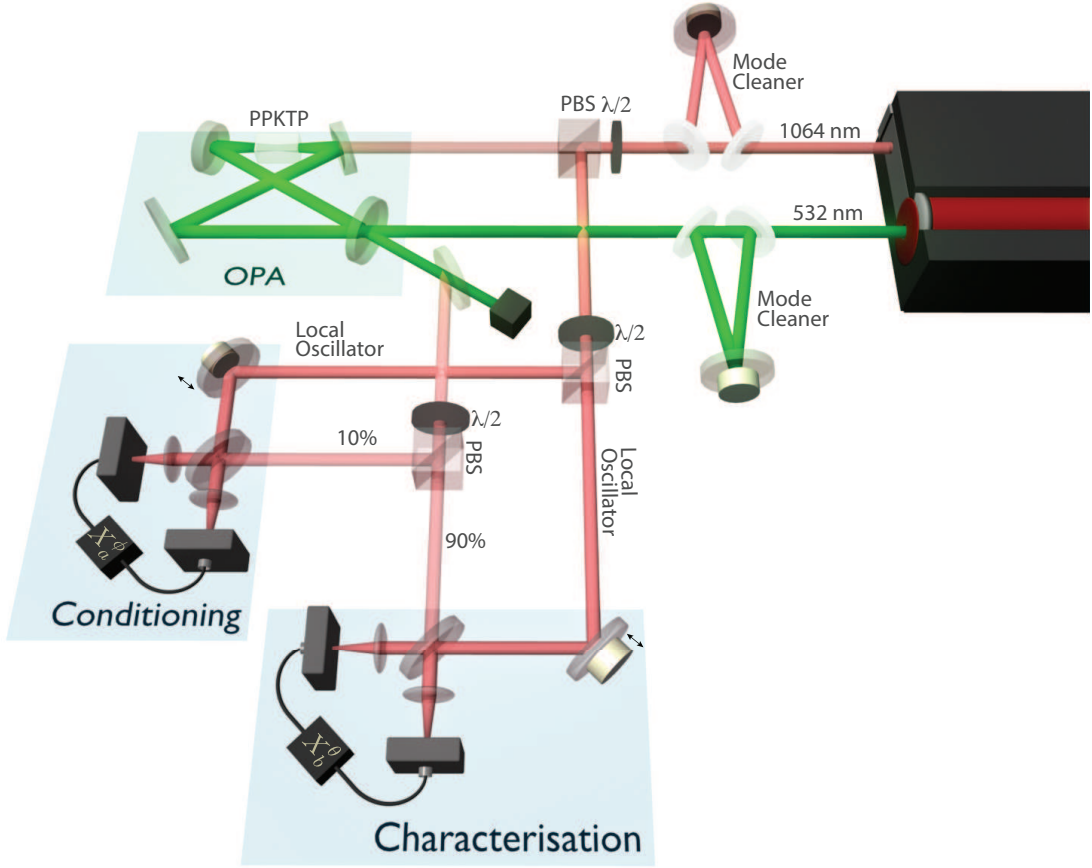}
\caption{
{\bf Experimental Setup} A CW Nd:YAG laser at 1064 nm provides the laser resource for this experiment. An internal second harmonic generation (SHG) cavity frequency doubles a portion of the 1064nm light. Both the 1064 nm and 532 nm fields undergo spatial and frequency filtering before providing seed and pump resources respectively for a doubly-resonant optical parametric amplifier (OPA). A small portion of the resulting squeezed coherent state is then reflected for 'conditioning' by a variable beam-splitter - implement with a $\lambda /2$ wave-plate and a polarising beam splitter (PBS). The reflected light (mode $a$) is subsequently sampled via a phase randomised homodyne detection. The remaining transmitted light (mode $b$) is characterised by a tomographic homodyne detection, sampling $X^{\theta}_{b}$ for $\theta = 0 \ldots 165 ^{\circ}$ in intervals of $15^{\circ}$. }
\label{figure2}
\end{figure*}

If $\rho_{ab}$ originates from a squeezed vacuum mode passing through
a weakly reflecting beam splitter, the resulting mode at $b$
conditioned on finding $n$ photons at $a$ will be an $n$-PSSV state.

We demonstrate how this conditioning can be performed using two
approaches. In the first approach, we express a polynomial operator
function of $\hat{n}_a$ that we want to condition upon in terms of
homodyne observables $\hat X^\phi_a$. In the second approach, we
utilise the pattern functions \cite{Leonhardt:1995ud} to access an
inner product in the Fock basis via homodyne measurements.

\subsection{Transformation Polynomials}
In this section, our goal is to obtain the measurement
statistics that would correspond to measuring a two mode observable $F(\hat
n_a)\otimes G(\hat{b})$ without actually constructing a device that
directly detects $F(\hat{n}_a)$. Instead we will be measuring quadrature
values of $a$. As we shall see in the following subsections, by suitably
conditioning on a homodyne measurement outcome of $a$, one can recover
the statistics of an $m$-photon subtracted state at $b$.

\subsubsection*{Example 1: Conditioning on $\hat{n}_a$.}

In the first example, we will attempt to condition the output state $b$ on the
measurement outcome of operator $\hat{n}_a$. We want to estimate the expectation value
\beq
\label{eq16}
g\left(X_b^\theta\right)=\trb{\rho_{ab} \;\hat{n}_a\otimes \ketbra{X_b^\theta}{X_b^\theta} }\;,
\eeq
where $\rho_{ab}$ is the joint state at modes $a$ and $b$. Expanding the operator $\hat n_a$, $g\left(X_b^\theta\right)$ can be
written as
\begin{align}
g\left(X_b^\theta\right) &= \sum_n n \,\mathrm{pr}(n) \trb{\rho_b(n)
  \ketbra{X_b^\theta}{X_b^\theta}}\\
&= \sum_n n\, \mathrm{pr}(n)\, \mathrm{pr}(X_b^\theta|n)\\
&= \sum_n n\, \mathrm{pr}(X_b^\theta,n)\;.
\end{align}
We use $\mathrm{pr}$ to denote probabilities and $\mathrm{pr}(n)$ denotes the probability
of getting an outcome $n$ at $a$. $\rho_b(n)$ is the state at $b$
conditioned on an outcome $n$ at $a$.

In particular, we consider $\rho_{ab}$ is a weakly squeezed vacuum state passing through a
low reflectivity beam-splitter with vacuum entering through the other input. Ignoring higher order terms, a squeezed
state can be approximated by $\ket{\psi}= \ket{0}-\ket{2}\gamma$ where $\gamma\ll
1$. $\ket{n}$ is the Fock state with $n$ photons. The beam-splitter transforms this state to
\beq
\ket{0,0}+ \left(\ket{1,1} \sqrt{2 \eta^2 (1-\eta^2)} +\ket{2,0} (1-\eta^2)+\ket{0,2}\eta^2 \right)\gamma
\eeq
where the beam splitter transmissivity is $\eta \sim 1$. $\ket{n,m}$
is the Fock state with $n$ photons in the first output (mode $a$) and $m$ photons
in the second output (mode $b$).

For this state, the expectation value, equation (\ref{eq16}) becomes
\beq
\label{eq21}
\trb{\left(\ketbra{1}{1} 2\eta^2(1-\eta^2) + \ketbra{0}{0}2(1-\eta^2)^2\right)  \ketbra{X_b^\theta}{X_b^\theta} }\gamma^2\;,
\eeq
where the second term arises from the probability of reflecting two photons. Assuming this probability is small (for $\eta \sim 1$), the output expectation value gives the statistics corresponding to a single photon Fock state.

To realise this conditioning, we could measure $a$ in the Fock basis
$\ketbra{n}{n}$ and scale the measurement outcomes of $b$ by the
outcomes $n_{a}$ (see Figure \ref{fig:con} (a)). But suppose we are
restricted to only homodyne tomography. We can still realise the
conditioning by expressing $\hat n$ in terms the quadrature
operators $\hat X$ and $\hat P $:
\beq
\label{eq22}
\hat{n} = \frac{1}{4}\left(\hat{X}^2+ \hat{P}^2 -2 \right)\;
\eeq
where $\hat X$ and $\hat P$ are two orthogonal quadrature operators
with the commutation relation $[\hat X,\hat P]=2 i$. Although $\hat X$ and
$\hat P$ cannot be measured simultaneously, Equation (\ref{eq16}) can nevertheless can be
written as the sum
\begin{align}
g\left(X_b^\theta\right) &=
\trb{\rho_{ab}\;\frac{1}{4}(\hat{X}^2-1) \otimes
  \ketbra{X_b^\theta}{X_b^\theta} } \\
&+\trb{\rho_{ab}\;\frac{1}{4}(\hat{P}^2-1) \otimes \ketbra{X_b^\theta}{X_b^\theta} } \;.
\end{align}
The expectation value $g\left(X_b^\theta\right)$ can be built up by
combining the outcomes of two independent measurements.

\underline{Phase randomised measurements}: The quadratures $\hat X$
and $\hat P$ can be replaced by any pair of orthogonal
quadratures. Instead of locking the quadrature angles, we can also
randomise the phase by scanning the local oscillator. Equation
(\ref{eq22}) can be written as an integration over all phases
\begin{align}
\hat{n} &= \frac{1}{2\pi} \int_0^{2\pi} \frac{1}{4} \left[
  (\hat{X}^\phi)^2 + (\hat{X}^{\phi+\frac{\pi}{2}})^2 -2\right]
d\phi\\
&=\frac{1}{2}\left( \bar{X}^2-1\right)\;,
\end{align}
where
\beq
\bar{X}^n = \frac{1}{2\pi} \int_0^{2 \pi}   (\hat{X}^\phi)^n d \phi
\eeq
is the phase averaged quadrature moment operator. Substituting this into Equation (\ref{eq16}) we obtain
\beq
g\left(X_b^\theta\right)=\trb{\rho_{ab}\; \frac{1}{2}\left( \bar{X}_a^2-1\right) \otimes \ketbra{X_b^\theta}{X_b^\theta} }\;.
\eeq
So $g\left(X_b^\theta\right)$ can be obtained by a phase randomised
sampling of the quadratures and weighting the outcomes at $b$ by the
outcomes of $\frac{1}{2}(\bar{X}_a^2-1)$ at $a$ (Figure \ref{fig:con}b).

\subsubsection*{Example 2: Conditioning on $\hat{n}_a(\hat n_a-2)$.}
To obtain a more faithful reproduction of the single photon Fock state
distributions from a weakly squeezed state, we can weight the outcomes on
$\hat{n}_a(\hat n_a -2)$ instead. This removes the contribution of two photon states at mode $a$. For a weakly squeezed vacuum input state (neglecting four photon terms), the analogue of Equation (\ref{eq21}) for this conditioning is
\beq
\label{eq15}
g\left(X_b^\theta\right)=\trb{ \ketbra{1}{1} 2\eta^2(1-\eta^2) \ketbra{X_b^\theta}{X_b^\theta}
}\gamma^2\;.
\eeq
To achieve this conditioning via homodyne measurements, we repeat the
recipe as before to express $\hat{n}$ in terms quadrature variables
$\hat X$ and $\hat P$:
\begin{align}
  \hat{n}(\hat n -2) &= \frac{1}{16} \left( \hat{X}^2 + \hat{P}^2
    -2\right) \left( \hat{X}^2 + \hat{P}^2 -10\right)\\
&=\frac{1}{16}\left(2\bar{X}^4 -24 \bar{X}^2+20 +\hat X^2 \hat
  P^2+\hat{P}^2 \hat{X}^2 \right)\;.\label{eq31}
\end{align}
The terms involving products of $\hat X$ and $\hat P$ cannot be evaluated directly through a phase randomised homodyne measurement. In order to make them accessible, we need to express $\hat X^2 \hat P^2+\hat{P}^2 \hat{X}^2 $  as a function of $\bar{X}$ which can be
  done as follows:
  \begin{align}
&    \hat X^2 \hat  P^2+\hat{P}^2 \hat{X}^2= \frac{1}{2\pi} \int
    _0^{2\pi} 2(\hat{X}^\phi)^2 (\hat{X}^{\phi+\frac{\pi}{2}})^2 d\phi\\
&=\frac{1}{\pi} \int_0^{2\pi} (2 \hat{a}_\phi^\dagger \hat{a}_\phi  \hat{a}_\phi^\dagger
\hat{a}_\phi +2 \hat{a}_\phi^\dagger \hat{a}_\phi -1- \hat{a}_\phi^4 - (\hat{a}_\phi^\dagger)^4)
d\phi\\
&=\frac{1}{\pi} \int_0^{2\pi} (2 \hat{a}_\phi^\dagger \hat{a}_\phi  \hat{a}_\phi^\dagger
\hat{a}_\phi +2 \hat{a}_\phi^\dagger \hat{a}_\phi -1 )d\phi\\
&=\frac{1}{\pi} \int_0^{2\pi} \left(\frac{( \hat{a}_\phi^\dagger
    +\hat{a}_\phi)^4}{3} -2\right) d\phi\\
&=\frac{2 \bar{X}^4}{3} -4\;,
  \end{align}
  where we define $\hat a_\phi = \hat a \exp(-i \phi)$. Substituting
  this into Equation (\ref{eq31}), we obtain the sampling polynomial as
\begin{align}
 \hat{n} (\hat n -2) = \frac{ \bar{X}^4}{6} -\frac{3 \bar{X}^2}{2}+1\;.
\end{align}
With this, the expectation value becomes
\beq
g\left(X_b^\theta\right)=\trb{\rho_{ab} \left( \frac{\bar{X}_a^4}{6} -
    \frac{3 \bar{X}_a^2}{2} +1\right) \otimes \ketbra{X_b^\theta}{X_b^\theta} }\;
\eeq
which can be sampled via a randomised phase quadrature measurement.

\subsubsection*{General conditioning on $f(\hat{n}_a)$.}
Higher order polynomials of $\hat n$ can be constructed in a similar
way. We provide two algorithms in Appendix A. These
polynomials provide a simple construction for a $k$ photon subtracted
state by conditioning on
\beq
\mathcal P(\hat{n})=\frac{1}{\hat{n}-k} \prod_{j=0}^{j_{max}} \hat{n}-j\;
\eeq
with $j_{max}>k$. Increasing $j_{max}$ in the product above would
correct for higher photon number contributions up to $j_{max}$. But
this will be at the expense of a higher weighting from outcomes having
photon numbers greater than $j_{max}$. If the probabilities of these
outcomes are large, it could dilute the actual conditioning state
that we are interested in.

As an example, to get a two photon subtracted state, we can use the
conditioning polynomial with $k=2$ and $j_{max}=6$:
\beq
\mathcal P(\hat{n})=\hat n(\hat n-1)(\hat n-3)(\hat n-4)(\hat n-5)\;.
\eeq
Expanding in the Fock basis,
\beq
\mathcal P(\hat{n}) = -12 \ketbra{2}{2} + 180 \ketbra{7}{7} + 1008 \ketbra{8}{8} + \ldots
\eeq
In this example, we see that the seven and eight photons events are weighted by a factor of 15 and 84 compared to the two
photon events. In most applications however, these high photon number states would have exponentially vanishing probabilities.

\subsection{Pattern functions}
The pattern functions, first introduced in~\cite{Leonhardt:1995ud,Leonhardt:1996wf}, specify the link between homodyne observables of a quantum state and the density matrix. These set of sampling functions allow reconstruction of the density matrix without the requirement of first reconstructing the Wigner function.

We want to characterise the state at $a$ conditioned on an $n$ photon event at $b$. Ideally, we would choose an appropriate polynomial in $X_{a}^{\phi}$ that corresponds to $\ketbra{n}{n}$. Practically, however, we can only realise a polynomial of a limited order---correcting for the finite undesired photon number events that may prove statistically
significant. The pattern functions however permit a perfectly isolating characterisation that removes all unwanted photon number events.

We start with the general problem of reconstructing the statistics of the post-selected state at $b$, $\tilde{\rho}_b$,
conditioned on the event of having a state $\rho_a^{cond}$ at $a$. This conditioning can be achieved by means of a measurement apparatus at $a$ having two outcomes:
\begin{eqnarray}
\pi_1 &= \rho_a^{cond} \\
\pi_2 & = 1- \rho_a^{cond}.
\end{eqnarray}
The output at $b$ conditioned on the outcome $\pi_1$ would be
\beq
\tilde{\rho}_b =\frac{1}{\mathrm{pr}_1} \ptrb{a}{\rho_{ab}\, \pi_1}
\eeq
where 
\beq
\mathrm{pr}_1 = \trb{\rho_{ab}\, \pi_1}
\eeq
is the probability of getting outcome $\pi_1$. We decompose the conditioned state $\rho_a^{cond}$ in the Fock basis with coefficients $c_{mn}$
\beq
\rho_a^{cond} = \sum_{mn} c_{mn} \ketbra{n_a}{m_a}
\eeq
so that the post-selected state at $b$ can be written as the sum
\begin{align}
\tilde{\rho}_b =\frac{1}{\mathrm{pr}_1} \sum_{mn}  c_{mn} \ptrb{a}{ \rho_{ab} \, \ketbra{n_a}{m_a}}\;.
\end{align}
To be able to reconstruct the post-selected state, we do a quadrature
tomography by measuring $X_b^{\theta}$ at $b$. The probability of
getting an outcome $X_b^{\theta}$ on the post-selected state is
\begin{align}
\tilde{\mathrm{pr}}\left(X_b^\theta\right) &=\bra{X_b^\theta} \tilde{\rho}_b\ket{X_b^\theta}\\
&=\frac{1}{\mathrm{pr}_1} \sum_{mn} c_{mn} \bra{m_a, X_b^\theta} \rho_{ab} \ket{n_a,X_b^\theta}\\
&=\frac{1}{\mathrm{pr}_1} \sum_{mn} c_{mn} \bra{m_a}\ptrb{b}{ \rho_{ab} \, \ketbra{X_b^\theta}{X_b^\theta}} \ket{n_a}\\
&=\frac{1}{\mathrm{pr}_1} \sum_{mn} c_{mn} \bra{m_a} \rho_{a}\left(X_b^\theta \right) \ket{n_a} \mathrm{pr}\left(X_b^\theta\right)\label{eq:pr_want}
\end{align}
where $\rho_{a}\left(X_b^\theta \right)$ is the state at $a$ when we
obtain outcome $X_b^\theta$ at $b$. The probability of getting this
outcome is denoted as $\mathrm{pr}\left(X_b^\theta\right)$.

We want to write the matrix elements $ \bra{m_a} \rho_{a}\left(X_b^\theta
\right) \ket{n_a} $ in terms of quadrature value measurements. For this
we utilise the Fock basis pattern function \cite{Leonhardt:1995ud} to
write
\beq
\bra{m_a} \rho_{a}\left(X_b^\theta \right) \ket{n_a} = \int_0^{\pi} \int_{-\infty}^{+\infty} \mathrm{pr}\left(X_a^\phi|X_b^\theta\right)
F_{mn}(X_a^\phi) d X_a d \phi
\eeq
where the $F_{mn}$ are the pattern functions of the Fock basis. They are given by
\beq
\label{patfun}
F_{mn}(X_a^\phi) = \frac{1}{\pi} \exp(i(m-n)\phi) \frac{\partial}{\partial x}\left[\psi_m(X_a) \varphi_n(X_a) \right]
\eeq
where $\psi_m(X_a)$ and $\varphi_m(X_a)$ are the $m$-th regular and irregular eigenfunctions of the Schr\"{o}dinger equation in a harmonic potential.
Substituting this into eq.(\ref{eq:pr_want}), we get
\begin{align}
\tilde{\mathrm{pr}}\left(X_b^\theta \right) &= \frac{1}{\mathrm{pr}_1}\sum_{mn}  c_{mn}  \int_0^{\pi} \int_{-\infty}^{+\infty}\mathrm{pr}\left(X_b^\theta \right) \mathrm{pr}\left(X_a^\phi|X_b^\theta\right) F_{mn}(X_a^\phi) d X_a d \phi \\
&= \frac{1}{\mathrm{pr}_1}\sum_{mn}  c_{mn}  \int_0^{\pi} \int_{-\infty}^{+\infty} \mathrm{pr}\left(X_a^\phi, X_b^\theta \right) F_{mn}(X_a^\phi) d X_a d \phi
\end{align}
where $\mathrm{pr}(X_a^\phi, X_b^\theta)$ is the unconditioned probability of
getting outcomes $X_a^\phi$ and $X_b^\theta$ when we measure $a$ and
$b$ in quadrature at angles $\phi$ and $\theta$.  Introducing the
weighting function
\beq
 w^{}\left( X_a^\phi\right) = \frac{1}{\mathrm{pr}_1}
\sum_{mn} c_{mn} F_{mn}\left( X_a^\phi\right)\;
\eeq
 we can write
\beq
\tilde{\mathrm{pr}}\left(X_b^\theta \right)= \int_0^{\pi}
\int_{-\infty}^{+\infty} \mathrm{pr}\left(X_a^\phi, X_b^\theta \right)
w^{}\left(X_a^\phi \right) d X_a d \phi\;.
\eeq
From this
expression, we see that the conditioned distribution
$\tilde{\mathrm{pr}}\left(X_b^\theta \right)$ can be obtained by sampling the
distribution $\mathrm{pr}\left(X_a^\phi,X_b^\theta \right)$ and weighting the
outcomes by $w^{}\left( X_a^\phi\right)$.

As an example, to obtain $\tilde{\rho}_b$ conditioned on a one photon
event at $a$, we condition on $\rho_a^{cond}=\ketbra{1}{1}$. This sets
$c_{11}=1$ and all other $c_{mn}=0$. To condition on the superposition
state $\rho_a^{cond}=\tfrac{1}{2}(\ket{1}+\ket{2})(\bra{1}+\bra{2})$, we require
$c_{00}=c_{01}=c_{10}=c_{11}=\tfrac{1}{2}$ and all other $c_{nm}=0$.

\underline{Phase randomised measurements}: For a conditioned state
$\rho_a^{cond}$ that is diagonal in the Fock basis, the weighting
function $w(X_a^\phi)$ is a sum of $F_{mn}(X_a^\phi)$ with $m=n$
which does not depend on the angle $\phi$. Hence the probability 
\beq
\tilde{\mathrm{pr}}\left(X_b^\theta \right)= \int_0^{\pi}
\int_{-\infty}^{+\infty} \mathrm{pr}\left(X_a^\phi, X_b^\theta \right)
w^{}\left(X_a \right) d X_a d \phi
\eeq
can be obtained by doing a phase randomised sample of the quadratures of $a$.

\section{Experiment}
\begin{figure*}
\includegraphics[width=15cm]{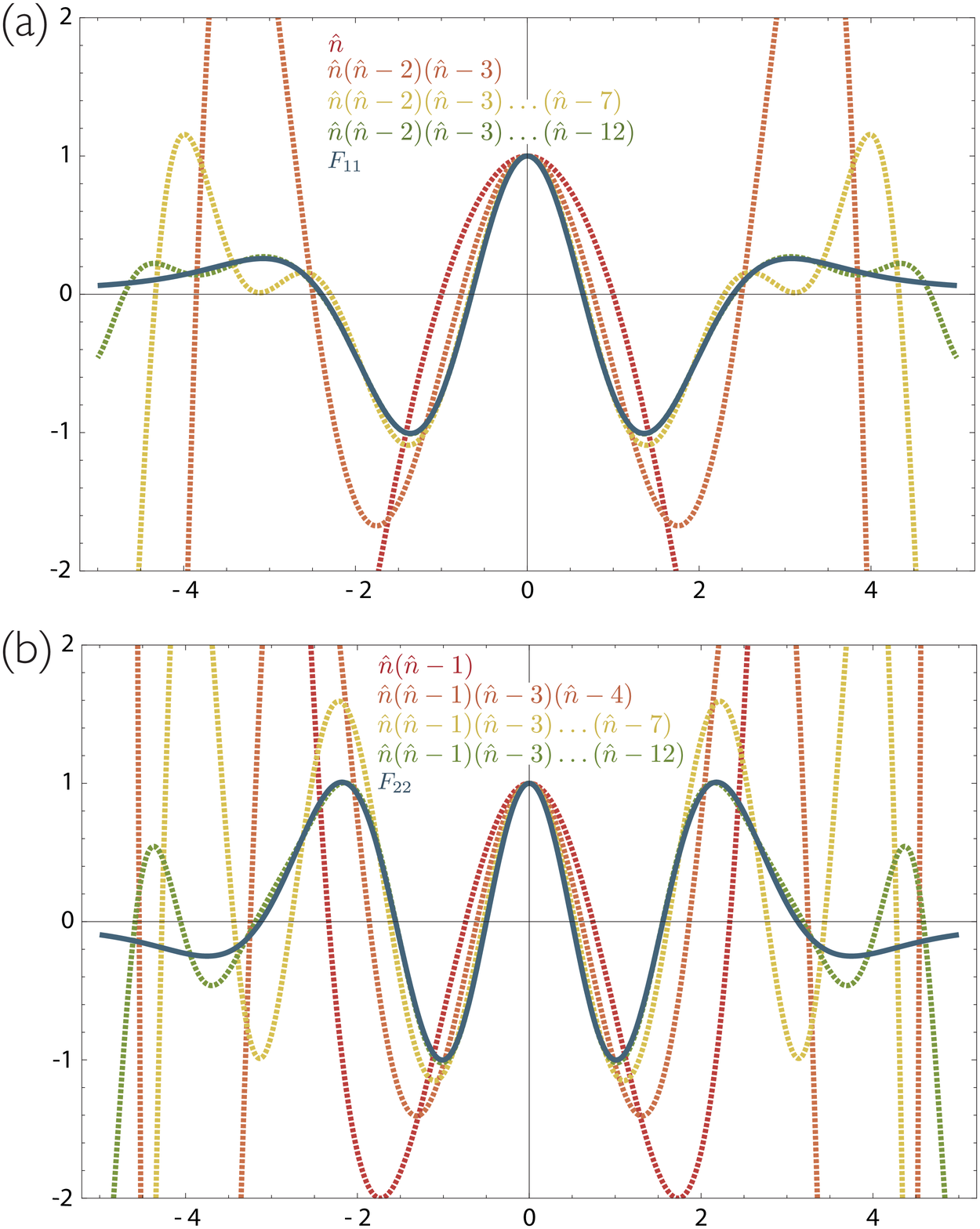}
\caption{
The convergence of the $\hat{n}$ polynomials to their corresponding
pattern functions for a photon number measurement of
(a) $n=1$ and (b) $n=2$. The polynomials are scaled so that
$\mathcal{P}(X)=1$ at $X=0$.}
\label{figure3}
\end{figure*}

\begin{figure*}
\includegraphics[width=\textwidth]{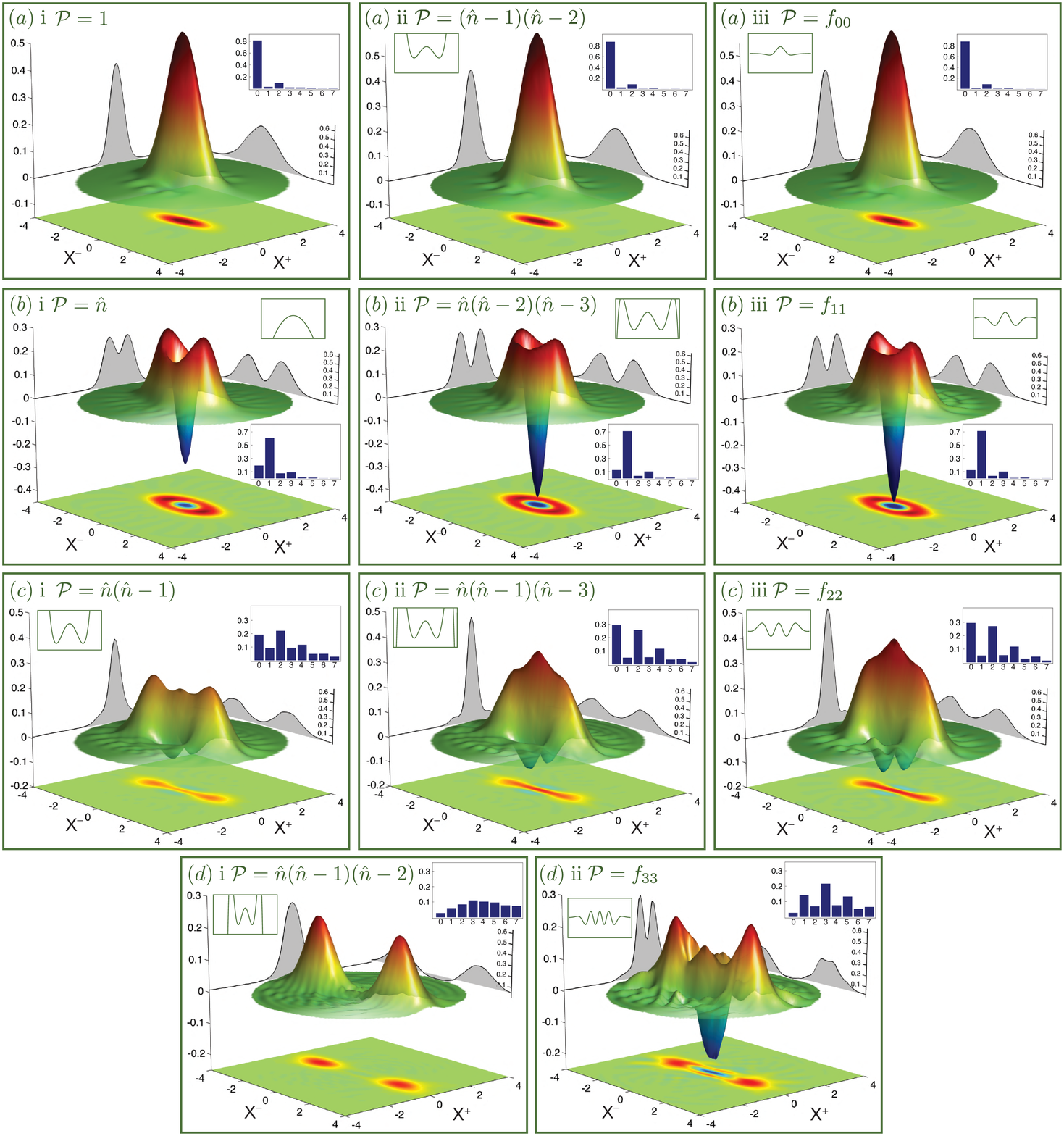}
\caption{ {\bf Reconstructed Wigner functions} of the: (a)
  $\mathrm{i}$ measured squeezed vacuum state and (a)$\mathrm{ii}$ and $\mathrm{iii}$ 
  $0$-PSSV state, (b) $1$-PSSV state, (c) $2$-PSSV state, and (d)
  $3$-PSSV state and their corresponding photon number populations (up to $n=7$),
  obtained from the reconstructed density matrix. Insets show the conditioning functions for each reconstruction. The Wigner functions are normalised such that the vacuum state has a variance of $\tfrac{1}{4}$.  }
\label{figure4}
\end{figure*}

Our experimental setup is detailed in Fig.\ref{figure2}. A
shot-noise limited 1064 nm Nd:YAG continuous wave (CW) laser provides
the laser source for this experiment. A portion of the 1064 nm light is
frequency doubled to provide a pump field at 532 nm. Both fields
undergo spatial and frequency filtering to provide shot-noise
limited light at the sideband frequencies above 2 MHz. A
doubly-resonant optical parametric amplifier in a bow-tie geometry
provides a squeezed vacuum resource at the sidebands centred around
the carrier. The resulting squeezed coherent state is then displaced to remove much of the intensity of the carrier field. The interference also provides classical amplitude and phase modulation signals, which was introduced on the reference beam, to control of the tomographic angle. The resulting dim squeezed state is then split,
with typically 10\% reflected towards the conditioning stage and subsequently sampled via phase
randomised homodyne detection.  The remaining transmitted 90\% is
measured via tomographic homodyne detection, consisting of sampling
$X_{b}^{\theta}$ for 12 values from $\theta = 0 \ldots 165^{
  \circ}$. The proportion of light reflected for conditioning is
sometimes increased to 15\% or 20\% to make unlikely events more
statistically accessible. This is done at the expense of the state
fidelity. Accurate state reconstruction with the techniques presented
here relies on experimentally realising a phase-randomised homodyne
detection with equal representation of all angles. This is
experimentally realised by sweeping the phase of the homodyne over a
few $\pi$ at approximately 100 Hz---significantly faster than the
drift of the global phase of the lasers. The encoding of phase and
amplitude modulation sidebands to allow control of the homodyne angle
$X^{\theta}_{b}$ for tomographic reconstruction also allows us to
verify that our conditioning measurement $X^{\phi}_{a}$ is
appropriately phase-randomised.

These two homodyne detections sample the sideband frequencies between
3--5 MHz, collecting typically $10^8$ samples per homodyne angle for
both characterisation and conditioning. The probability distributions
for the directly measured squeezed state $X^{\theta}_{b}$ is
reconstructed for the 12 measured values of $\theta$. We then employ a
maximum entropy state estimation (as originally defined in
\cite{Buzek:2000hy}) permitting a Hilbert space up to $n$=30.  The
maximum entropy state estimation gives the most mixed state consistent
with our measured statistical ensemble.

\section{Results}

If we first ignore the role of conditioning, the ensemble of homodyne
measurements at the tomographic characterisation stage allows
construction of the histograms describing the probability distribution
of each measured $X^{\theta}_{b}$. To achieve this, for each sample,
$X^{\theta}_{b}$, we increment the relevant bin by one. One can
then reconstruct the Wigner function of the state sampled at the
tomographic homodyne detection (Figure \ref{figure4}(b)i) using the
maximum entropy state estimation principle \cite{Buzek:2000hy}. The extension to
`conditioning' in post-processing is implemented as follows. For each
sample $X^{\theta}_{b}$ we have a corresponding measurement of mode
$a$, $X^{\phi}_{a}$, which provides the value for the relevant
weighting. Instead of incrementing the bin corresponding to
$X^{\theta}_{b}$ by one, we increment the bin by the outcome
of a function of our choosing $\mathcal{P}(X_a^{\phi})$.

In Figure \ref{figure4} (b) we focus on the reconstruction of the
1-PSSV state.  Figure \ref{figure4} (a) $\mathrm{i}$ shows the Wigner
function obtained using the simplest conditioning polynomial,
$\mathcal{P}(X^{\phi}_{a}) = \hat{n}_{a}$. This should ideally remove
any contributions corresponding to a measurement of $n_a = 0$ (vacuum)
in mode $a$. All other contributions remain and their contributions
are additionally weighted by their corresponding eigenvalues,
$n_a$. In essence we reconstruct a statistical mixture of primarily
the $1$-PSSV and $2$-PSSV states, where their contributions are not
solely weighted by the likelihood of successful `conditioning', but
additionally by their corresponding eigenvalue. For instance, the
contributions from $n_a =2$ are weighted at twice that of
contributions from $n_a = 1$. 

An idealised implementation of a photon annihilation corresponds to a beam splitter with reflectivity
approaching zero.  This permits statistical isolation of a single
photon subtraction event from the considerably less likely two photon
subtraction event. However, with an experimental implementation, the
requirement of a finite tap-off (typically around 10\%) inevitably
introduces spurious higher order photon subtraction contributions.

One can instead consider a higher order polynomial in $\hat{n}_a$ that
removes potential contributions to the reconstructed state from higher
order subtractions that are unwanted and are sufficiently
statistically significant to warrant removal. As the ideal squeezed vacuum populates only the even photon number pairs, the ideal subtraction of one photon from squeezed vacuum should produce a superposition of the odd photons numbers (and remove any vacuum contribution). Figure \ref{figure4} (b)
$\mathrm{ii}$ demonstrates the dramatic improvement in the
reconstructed $1$-PSSV state by implementing the conditioning
polynomial $\mathcal{P}(X^{\phi}_{a}) = \hat{n}_{a}(\hat{n}_{a} -2 )
(\hat{n}_{a} -3)$, removing polluting contributions from the $2$ and
$3$ photon subtractions.  The
$F_{11}$ pattern function allows an ideal implementation of a one
photon conditioning in mode $a$ (Figure \ref{figure4}
(b)$\mathrm{iii}$). The results of Figure \ref{figure4} (b)
$\mathrm{ii}$ and $\mathrm{iii}$ are markedly similar (sharing a fidelity of 99.2\%) despite the
clear departure of the polynomials, especially noting how rapidly the
polynomial $\hat{n}_{a}(\hat{n}_{a} -2 ) (\hat{n}_{a} -3)$ diverges in
$X^{\phi}_{a}$ (Figure \ref{figure3}).

Figure \ref{figure4} (c) compiles the results of the $2$-PSSV state
reconstruction. Figure \ref{figure4} (c) $\mathrm{i}$ considers the
$\mathcal{P}(X^{\phi}_{a}) = \hat{n}_{a}(\hat{n}_{a} -1) $, removing
contributions corresponding to a photon number measurement of $n_a =
0$ and $n_a = 1$. The ideal reconstructed $2$-PSSV state has high
fidelity with the even kitten state. When we additionally correct for
the contributions of the $3$-PSSV state there is a clear improvement
(Figure \ref{figure4} (c) $\mathrm{ii}$) in the purity of the
reconstructed state, evidenced by the increasing isolation of the
even-photon number contributions to the photon number populations. If
we consider the relevant pattern function $\mathcal{P}(X^{\phi}_{a}) =
F_{22}$ (Figure \ref{figure4} (c) $\mathrm{iii}$) we see a further
improvement in the purity of the reconstructed state.

There a handful of subtleties involved in the estimation of the photon
statistics with homodyne measurements. Analogies with many of these
can be drawn with the usual problems that afflict photon counting
measurements. This technique relies on correlations shared between
modes $a$ and $b$, and may be degraded by any process that introduces
uncorrelated classical or quantum noise. With the results presented
here, the significance of electronic noise in detection is understood
to be negligible. As we adopt an inherently `ensemble' approach, by
making the assumption that the dark noise is uncorrelated to the
quantum state we could realise a dark noise correction for both our
conditioning in mode $a$ and our characterisation in mode $b$. In
reality, the dark noise is sufficiently negligible that any correction
proves insignificant. Experimentally, we typically enjoy greater than
18 db dark noise clearance over our measurement band.

However, we are still exposed to the effects of loss. Any loss of
purity on the initial squeezed vacuum state constrains the
non-Gaussian nature of the reconstructed state. The role of loss can
be accurately modelled as a beam-splitter with transmissivity
$\lambda$. The role of loss can be understood in by drawing analogy to
traditional photon counting. Inefficiencies arising from imperfect
homodyne detection efficiency or transmission losses scale the rate of
success of the homodyne conditioning, analogous to loss on a photon
counting measurement. Whilst here we cannot refer to individual
events, as this approach succeeds by considering the entire ensemble,
we essentially require a larger ensemble to obtain the same
conditioned statistics. Additionally, it can also lead to erroneous
conditioning, where a loss of photon may see a 3-photon subtraction
event contributing as two photon subtraction.

Our homodyne efficiency is typically 98\%, with a fringe visibility of typically $99.2\%$ and specified photodiode quantum
efficiency of $\geq 99\%$. Our primary source of loss in the
experiment arises from the impurity of the squeezed vacuum
resource---and this is most evident with the reconstruction of the
3-PSSV state (Figure \ref{figure4} (d)). Endeavouring to reconstruct
the 3-PSSV state, we optimised the experimental parameters to increase
the likelihood of having 3 photons in mode $a$ without sacrificing the
quality of the reconstructed state. The likelihood of encountering a 3
photon subtraction event is low. Whilst the probability of subtracting
$n$-photons with a beam splitter of reflectivity $\eta$ scales as
$\eta^{n}$, attempting to measure 3 or 4 photons from mode $a$ also
enforces the additional requirement of having at least $4$ photons in
the original squeezed vacuum mode. As a result, the likelihood of
having 3 or more photons in mode $a$ scales poorly. We can improve
this predicament by firstly increasing the percentage of the input
mode used for conditioning (typically 15\%) and secondly, by using a
stronger squeezed resource, enhancing population of the higher order
photon pairs. Increasing the squeezing level is detrimental to the
squeezing purity as it introduces noise sources only dominant at high pump power, such as phase
noise. In our doubly-resonate system the requirement of the stronger
pump field also has consequences for the long-term stability of the
experiment. Obtaining sufficient statistics requires longer
acquisition time which concatenates the typical experimental drifts in
the measured tomographic angle $\theta$, alignment and squeezing
levels over time, reducing the overall purity of the reconstructed
state. As a result the reconstructed $3$-PSSV state in Figure
\ref{figure4} (d) has lower reconstructed state purity (evidenced by
the smaller observable negativities at the origin) than the
reconstructed $1$ and $2$-PSSV states which require smaller data sets.

If we attempt to reconstruct the $3$-PSSV state with an additional
correction for the $4$ photons events in mode $a$, the reconstructed
state becomes noisier. It is not immediately apparent that removing
unwanted contributions should introduce statistical noise into the
ensemble, but the conditioning on higher photon numbers or the removal
of higher order terms essentially requires extraction of finer
correlations between modes $a$ and $b$. For a polynomial
$\mathcal{P}(n_a)$ of degree $k$, we essentially estimate moments of
$X^{\phi}_{b}$ up to order $2k$. When coupled with the rapid
divergence of the polynomials in $X^{\phi}_{b}$, sufficient statistics
must be acquired to minimise error. This prevents us from implementing a purification of the $3$-PSSV state in figure \ref{figure4} (d) with the polynomial approach, even though it is successful with the corresponding $F_{33}$ pattern function (figure \ref{figure4} (d) $\mathrm{ii}$).

While the pattern functions extract the statistics of ideal photon
number discriminating measurement at mode $a$, limited only by the
experimental imperfections, it is worth noting that one can
essentially obtain the same outcome by implementing a polynomial
weighting to only a few orders. This is despite the fact the
polynomials calculated to any $\mathcal{P}(\hat{n_a})$ will diverge
for sufficiently large $X^{\theta}_{b}$. In spite of the clear
divergence between polynomial $\hat{n}_{a}(\hat{n}_{a} -1 )
(\hat{n}_{a} -3)$ and the corresponding pattern function $F_{22}$
(Figure \ref{figure3} (b)), the corresponding Wigner functions (Figure
\ref{figure4} (c) $\mathrm{ii}$ and $\mathrm{iii}$ share a fidelity of
$98.8$\%. To emulate a conditioning photon number measurement a low
order implementation of the $\hat{n}$ polynomials is generally
sufficient.

As a small aside, we also consider the effect of measuring no photons in the conditioning mode (Figure \ref{figure4} (a) {\rm ii} and {\rm iii}). This projects onto a subset of weaker squeezed vacuum input states. This can be compared to the action of de-amplification with a noiseless linear amplifier with a gain $< 1$.

\section{Conclusion}
In this paper we have experimentally demonstrated a technique to
reconstruct the Wigner functions of various non-Gaussian states of
light with only homodyne measurements. This technique relies on an
ensemble based post-processing of the homodyne data informed by a
phase randomised homodyne measurement. While it therefore never allows
us to prepare a non-Gaussian state, it still enables their
characterisation. Using these methods, we were able to reconstruct a
1-PSSV, 2-PSSV and 3-PSSV. Previously, extracting such statistics
would have required a full tomographic reconstruction of the two-mode
Wigner function. These techniques allow for complete characterisation
of the outcome of a conditional measurement on a system, and might
prove useful in systems where measurements of the DV of the system are
limited or unavailable.

\begin{center}\begin{tabular*}{.5\textwidth}{c}\hline\end{tabular*}\end{center}

This research was conducted by the Australian Research Council Centre of Excellence for Quantum Computation and Communication Technology (Project number CE110001027).

\appendix
\section{Conditioning Polynomials}
In this appendix, we demonstrate how the sampling polynomials can be
obtained for arbitrary functions of $\hat n$. We provide two
equivalent methods for doing this.

The first method involves writing the polynomial functions of the
phase randomised quadrature operators $\bar{X}$ in terms of $\hat{n}$ via
the creation and annihilation operators. These functions can then be
inverted to solve for functions of $\hat n$ in term of $\bar{X}$. 

The second method reproduce the same polynomials via measuring the
moment of the Fock state by integration of Hermite polynomials.

\subsection*{Method 1}
For an arbitrary function of $f(\hat{n})$, the analogue of equation (\ref{eq16})
that we want to estimate using a phase randomised homodyne measurement
would be
\begin{align}
f\left(X_b^\theta\right)&=\trb{\hat \rho_{ab} f(\hat{n})\otimes
  \ketbra{X_b^\theta}{X_b^\theta} }\\
&=\mathrm{pr}(X_b^\theta)\ptrb{a}{\hat \rho_{a}(X_b^\theta) f(\hat{n})}\;,
\end{align}
where $\rho_a(X_b^\theta)$ is the state at $a$ after tracing out
$b$. Our goal is to find a function $F(\bar{X})$ corresponding to
$f(\hat n)$ such that
\beq
\label{eqA3}
\trb{\hat{\rho} f(\hat{n})} = \trb{\hat{\rho} F(\bar X)}\;,
\eeq
where
\beq
\label{eqA4}
F(\bar X) = \frac{1}{2\pi} \int_0^{2\pi}\, d\theta F(\hat a_\phi + \hat a_\phi^\dagger)
\eeq
and $\hat a_\phi =\hat a \exp(-i \phi)$. Let us consider polynomial
functions of $\bar{X}$ for which the monomials $\bar{X}^m$ for
$m=0,1,\ldots$ forms a basis.

For all odd values of $m$, $\bar{X}^m$ vanish since the exponential terms $\exp(-i \phi)$
integrate to zero. For even $m$, the only terms in the expansion of
$(\hat a_\phi+\hat a_\phi^\dagger)^m$ that are not a function of $\phi$ are those
having equal numbers of $\hat a_\phi$ and $\hat a_\phi^\dagger$. These are the only terms that are non-zero after performing the integral in equation (\ref{eqA4}). They can be expressed as a function of $\hat n$ using the identity $\hat a^\dagger \hat a=\hat{n}$ and the commutation relation $[\hat a,\hat a^\dagger]=1$.

We provide an example for the case of $m=4$:
\begin{align}
  \bar{X}^4 &= \frac{1}{2\pi} \int_0^{2\pi} d\phi \left(\hat{a}_\phi +
  \hat{a}_\phi^\dagger\right)^4\\
&=\hat{a}\hat{a}\hat{a}^\dagger\hat{a}^\dagger+\hat{a}\hat{a}^\dagger\hat{a}\hat{a}^\dagger+\hat{a}\hat{a}^\dagger\hat{a}^\dagger\hat{a}\\
&+\hat{a}^\dagger\hat{a}^\dagger\hat{a}\hat{a}+\hat{a}^\dagger\hat{a}\hat{a}^\dagger\hat{a}+\hat{a}^\dagger\hat{a}\hat{a}\hat{a}^\dagger\\
&=6\hat{n}^2 + 6\hat{n} +3\;.\label{eqA5}
\end{align}
Results for various powers of $\bar{X}$ are tabulated below.
\begin{align*}
\bar{X}^0 &= 1\\
\bar{X}^2 &= 1+2\hat{n}\\
\bar{X}^4 &= 3+6\hat{n} + 6 \hat{n}^2\\
\bar{X}^6 &= 15+40\hat{n} + 30 \hat{n}^2 + 20 \hat{n}^3\\
\bar{X}^8 &= 105 +280\hat{n} + 350 \hat{n}^2 + 140 \hat{n}^3+ 70 \hat{n}^4\\
\bar{X}^{10} &= 945+ 2898\hat{n} + 3150 \hat{n}^2+ 2520 \hat{n}^3+ 630 \hat{n}^4+ 252 \hat{n}^5
\end{align*}

\subsection*{Method 2}
As an alternative method, we note that equation (\ref{eqA3}) must hold for arbitrary inputs $\hat{\rho}$. In
particular, when $\hat{\rho}=\ketbra{n}{n}$ we get
\begin{align}
  \trb{F(\bar{X})\ketbra{n}{n} } &= f(n)\\
\iint dx d\tilde{x} \,\phi_n(x) \phi^\star_n(\tilde{x})F(x) \delta(x-\tilde{x}) &=f(n)\\
\int dx\, \left| \phi_n(x) \right|^2 F(x) &= f(n)
\end{align}
where $\phi_n(x) = \left< n|x \right>$  are the eigenstates of the harmonic
oscillators. For $F(\bar{X})=\bar{X}^m$, the associated functions of
$n$ would correspond to the $m$-th moment of the eigenstates.

While this integration can be performed directly using the
Hermite polynomials, it turns out that it is more convenient to
express $\bar{X}$ in terms of the annihilation and creation operators
instead. As an example, we evaluate $f(n)$ when
$F(\bar{X})=\bar{X}^4$:
\begin{align}
 \bra{n} \bar{X}^4 \ket{n} &= \frac{1}{2\pi} \int_0^{2\pi} d\phi \bra{n}\left(\hat{a}_\phi +
  \hat{a}_\phi^\dagger\right)^4 \ket{n}\\
&=\bra{n}\hat{a}\hat{a}\hat{a}^\dagger\hat{a}^\dagger+\hat{a}\hat{a}^\dagger\hat{a}\hat{a}^\dagger+\hat{a}\hat{a}^\dagger\hat{a}^\dagger\hat{a}\\
&+\hat{a}^\dagger\hat{a}^\dagger\hat{a}\hat{a}+\hat{a}^\dagger\hat{a}\hat{a}^\dagger\hat{a}+\hat{a}^\dagger\hat{a}\hat{a}\hat{a}^\dagger\ket{n}\\
&=6n^2 + 6n +3\;
\end{align}
which is the same result as equation (\ref{eqA5}) as to be expected.

\end{document}